\documentclass[review,12pt]{elsarticle}
\usepackage{geometry}
\usepackage{amssymb}
\usepackage{amsmath}
\usepackage{hyperref}
\hypersetup{hidelinks}
\usepackage{booktabs}

\begin{document}

\begin{frontmatter}

\title{A Unified Theoretical Treatment on Statistical Properties of the Semi-batch Self-condensing Vinyl Polymerization System}

\author[1]{Fang Gu}
\author[1]{Jiangtao Li}
\author[2]{Xiaozhong Hong\corref{corresponding author B}}
\cortext[corresponding author B]{Corresponding author}
\ead{hxz@hbu.edu.cn}
\author[1,3,4]{Haijun Wang\corref{corresponding author}}
\cortext[corresponding author]{Corresponding author}
\ead{whj@hbu.edu.cn}

\address[1]{College of Chemistry and Environment Science, Hebei University, Baoding, 071002, Hebei, China}
\address[2]{College of Physical Science and Technology, Hebei University, Baoding, 071002, Hebei, China}
\address[3]{Chemical Biology Key Laboratory of Hebei Province, Hebei University, Baoding, 071002, Hebei, China}
\address[4]{Key Laboratory of Medicinal Chemistry and Molecular Diagnosis, Ministry of Education, Hebei University, Baoding, 071002, Hebei, China}

\begin{abstract}
We present a novel generating function (GF) method for the self-condensing vinyl polymerization (SCVP) system with any initial distribution of preexisted polymers. Such a method was proven to be especially useful to investigate the semi-batch SCVP system allowing a sequence of feeding operations during the polymerization.
Consequently, the number-, weight-, and z-average molecular weights as well as polydispersity index of hyperbranched polymers can be explicitly given, which are determined by predetermined feeding details and conversions in each polymerization step.
These analytical results are further confirmed by the corresponding Monte Carlo simulation.
Therefore the present GF method has provided a unified treatment on the semi-batch SCVP system. Accordingly, hyperbranched polymers with desired properties can be prepared by designing feeding details and presetting conversions at each step based on the present GF method.
\end{abstract}


\end{frontmatter}


\newpage

\section{Introduction}
Since the pioneering discovery of dendrimers due to Tomalia and coworkers\cite{Tomalia1986},
dendritic polymers such as dendrimers and hyperbranched polymers have caused particular interests for their special structures and properties\cite{Kim1990,Frechet1995,Gao2004,Voit2009}.
Compared with dendrimers,
hyperbranched polymers have imperfect structures,
but possess the features of relatively convenient synthesis and low cost\cite{Gao2004,Voit2009},
and therefore can be easily industrialized.
With the numerous relevant studies,
some characteristic properties of hyperbranched polymers have been found such as low viscosity,
high solubility, without entanglement, and with a large number of functional terminal groups.
So far,
hyperbranched polymers have been widely utilized in blending modification,
surface modification, coating, catalytic reaction, and drug delivery\cite{Gao2011,Zheng2015}.

As is well known, there are two representative methods for preparing hyperbranched polymers.
One is the polycondensation of $\mathrm{AB}_{g}$ ($g \geq 2$) type monomers
proposed initially by Flory\cite{Flory1952}, and came true by Webster and Kim \cite{Kim1990}.
The other is the self-condensing vinyl polymerization (SCVP) of AB* type inimers proposed by Fr\'{e}chet in 1995\cite{Frechet1995},
where A denotes a vinyl group and B* represents an active site capable of initiating
a vinyl group A into a new active site A*.
In the light of this synthetic route, a mass of experiments on several typical SCVP systems have been performed,
which include SCVP systems of pure inimers\cite{Podwika1997,Gaynor1997,Simon1997,Simon2004,Dong2010,Pugh2010},
binary copolymerization systems(inimers and core initiators\cite{Simon2001,
Pan2001,Hong2001,Georgi2010,Lu2017},
inimers and monomers\cite{Mori2002,Mori2004,Jutz2004,Mori2005,Liu2008,Aydogan2016}),
and ternary copolymerization system(inimers, monomers and core initiators) \cite{Mori2006,Xue2017,Aydogan2017}.
These experiments have manifested that these SCVP systems had an appreciable prospect not only in polymer science but also in industry and chemical engineering.

Following these relevant experiments\cite{Podwika1997,Gaynor1997,Simon1997,Simon2004,Dong2010,Pugh2010,
Simon2001,Pan2001,Hong2001,Georgi2010,Lu2017,Mori2002,Mori2004,Jutz2004,
Mori2005,Liu2008,Aydogan2016,Mori2006,Xue2017,Aydogan2017,Alfurhood2016},
some theoretical investigations on SCVP systems have been well presented.
In 1997, the SCVP system of pure inimers was investigated for
the first time by M\"{u}ller, Yan and Wulkow\cite{Wulkow1997,Yan1997}.
Subsequently, binary and ternary SCVP systems
have been extensively studied by various methods.\cite{Radke1998,
Yan1999,Litvinenko1999,Litvinenko2001,Litvinenko2002,Cheng2003,He2001,He2003,
Zhou2008,Zhou2009,Zhao2009,Zhao2011,Yan2008,Hong2015,Tobita2015,Tobita27,
Tobita28,Tobita2019}.
Among these works, much attention has been paid to the polymerization mechanism, kinetics, size distribution, average molecular weights, polydispersity, degree of branching, mean-square radius of gyration, and so forth.
These theoretical works have not only
offered rich information to learn about various SCVP
systems but also inspired some new research interests.

With the rapid developments in both theoretical and experimental studies on different SCVP systems,
the semi-batch process (known also as the semi-continuous process) has attracted an increasing interest since
it is closely related to large-scale applications of hyperbranched polymers in various fields\cite{Cheng2003B,Cheng2005,Zhu2006,Wang2010,Zhu2012,Zhu2013,Cheng2014,
Ilchev2015,Cheng2017}.
It is well known that for a polymerization system,
the procedure by which a polymerization is carried out can lead to a significant effect on the polymerization kinetics,
dynamics and physical properties of resultant polymers.
As far as this aspect is concerned,
depending
on the manner that the reactants are added into the reactor,
polymerization processes can be classified into the batch, semi-batch, and continuous processes, respectively.

In a batch process, all the reactants are fed into the reactor at the beginning of a polymerization,
and no material is added into or removed from the reactor during the polymerization.
A semi-batch process refers to the case that reactants would be added into the reactor during the polymerization, while the products may also be removed from the reactor.
In general, only a portion of reactants is
initially loaded into the reactor, whereas the remainder would be fed according to a predetermined schedule.
Unlike batch and semi-batch processes, a continuous process involves both a feeding of reactants and a removal of products such that there is usually a balance between the input and output streams.

In essence, these three types of feeding operations signify different local reaction conditions such as concentrations of individual species
including monomers, initiators, and possible chain transfer agents.
As such, polymers produced from the same reactants may
possess quite different properties depending on the distinct feeding processes.
It is therefore expected that there is an obvious effect of the feeding process on the final properties of polymers such as the average molecular weights, polydispersity, degree of branching, and possible copolymer composition.

Experimentally,
a batch process needs the modest demanding and the least amount of control,
and hence it is mainly used for pure academic interests.
A drawback in the batch process is that the drifts of molecular weight distribution and composition distribution are very common because reactant concentrations decrease gradually during the polymerization.
In contrast to the batch process,
a continuous process tends to be applied when a large volume of polymers is manufactured.
In practice,
most of continuous processes are performed at a relative steady reaction conditions.
In this way,
the drift of molecular weight distribution can be well inhibited.
However, it is inevitable that a high instrumentation cost and a long uninterrupted reaction time are usually involved.
By comparison,
in a semi-batch process,
the corresponding drifts can be substantially reduced since the concentration of reactants can be controlled by an appropriate predetermined feeding schedule.
Consequently,
a polymerization can even be carried out under starved conditions to prepare polymers with desired properties.

In addition to the above-mentioned features,
the versatility of the semi-batch mode is also closely related to the problem of exotherm.
As is well known, some living polymerizations are so fast that the
released heat due to polymerization becomes a potential hazard.
However, controlling the feeding details in the semi-batch process
allows a control over the reaction rate, and hence the exotherm.
In other words,
a low concentration of reactants can be maintained such that the semi-batch operation is compatible with the safe requirement.
Moreover,
the semi-batch process can also provide a regulation over the polymer structure and morphology for some polymerization systems.\cite{Cheng2003B,Cheng2005,Zhu2006,Wang2010,Zhu2012,Zhu2013,Cheng2014,
Ilchev2015,Cheng2017}
In view of these obvious advantages,
the semi-batch process has been widely utilized in both industry and academic interests.

This work is just motivated by the use of semi-batch process in the SCVP system consisting of pure inimers.
An attempt is made to find how the semi-batch process regulates the average properties of hyperbranched polymers.
This idea is based on the fact that a semi-batch process can always be decomposed into a series of time-sequenced feeding and polymerization steps.
More importantly, the polymerization occurred between any two nearest neighbor steps is in essence equivalent to a batch process with various preexisted polymers formed by another SCVP system.
Therefore such a goal can be achieved once an average polymeric
quantity associated with any two nearest neighbor steps is presented.
This signifies that,
if an average polymeric quantity for any two nearest neighbor polymerization steps is figured out,
one is able to find the effect of the semi-batch process on the SCVP system.

For this purpose,
our discussions will be firstly focused on the two nearest neighbor feedings, and then we generalize it to the whole feeding sequence.
As usual,
the number-, weight-, and z-average molecular weights denoted by $M_{\mathrm{n}},
M_{\mathrm{w}}$ and $M_{\mathrm{z}}$ as well as polydispersity index (PI)
would be investigated.
Although the semi-batch SCVP system is very complicated, these average physical quantities can still be obtained in a straightforward manner.
Throughout the paper the following classical assumptions of ideal polymerization were adopted:
(1). Two types of active sites B* and A* have the same equal reactivity;
(2). All the vinyl group A react independently one another;
(3). No intramolecular reaction takes place in the system.
These approximations have ever been used in previous studies.\cite{Wulkow1997,Yan1997,Radke1998,Yan1999,
Litvinenko1999,Litvinenko2001,Litvinenko2002}
For simplicity, the mass of an inimer has been taken as the unit mass.

This article is organized as follows.
In section 2, an explicit generating function (GF) is proposed to calculate some average properties of the SCVP system with any initial distribution of preexisted polymers.
Upon applying the GF method to the batch SCVP system of pure inimers,
it reproduces identical results as given by previous studies,
thereby validating the present GF method.
In section 3, we generalized the present GF method to the semi-batch SCVP system, and investigated $M_{\mathrm{n}}$, $M_{\mathrm{w}}$, $M_{\mathrm{z}}$, and PI of hyperbranched polymers.
This is based on the fact that an SCVP system with preexisted polymers corresponds to two nearest neighbor steps in the semi-batch SCVP system.
Meanwhile,
two typical feeding processes are taken to illustrate the effect of the semi-batch process on these average polymeric quantities.
In section 4,
a Monte Carlo (MC) simulation is performed for the semi-batch SCVP system to confirm the present GF method.
As expected, an excellent agreement between analytical and simulation results is observed.
In section 5,
we summarize some relevant results and discuss some limitations of the present work.

\section{Generating Function Method for an SCVP System with Any Initial Distribution of Polymers}

As stated above,
the effect of the semi-batch process on an SCVP system
can be presented as long as the relationship of an averaged polymeric quantity associated with any two nearest neighbor feedings is obtained.
In this section, a novel GF method will be proposed to achieve such a goal.
Without loss of generality,
we suppose that there have preexisted some polymers at the beginning of the an SCVP system under consideration.
This could correspond to two nearest neighbor polymerization steps, where the preexisted polymers come from either a previous polymerization step or a feeding operation, or from both them.
In this way, an explicit GF will be employed to study such an SCVP system, which enables us to find how an average polymeric quantity
depends on these preexisted polymers.

To proceed,
we suppose that, for the present SCVP system,
the number of preexisted polymers of $m$-mer ($m=1,2,3,...$) in the initial stage is $P_{m}(x=0)$,
where the symbol $x$ denotes the conversion of vinyl groups A.
According to the reaction mechanism of the SCVP\cite{Frechet1995,Wulkow1997,Yan1997},
an active site B* or A* can initiate a vinyl group A of another molecule to
form a new molecule and generate a new active site A*.
Clearly, the consumption of an old active center (A*or B*) always leads to a new active center A*.
This means that the number of active sites remains unchanged during the polymerization.
It is also evident that there exist $m$ active sites and only one vinyl group A in an $m$-mer in accordance with the assumption of neglecting intramolecular reaction.
Thus the total number of active sites $N_{\ast}$ is $N_{\ast}=\sum_{m}mP_{m}(x=0)$,
while the total number of vinyl groups A is $N=\sum_{m}P_{m}(x=0)$.
This is the initial condition of the present SCVP system.

Through successive reactions,
molecules in the system can be linked together to form hyperbranched polymers of various sizes.
As proposed by M\"{u}ller, Yan and Wulkow, the evolution of the size distribution of polymers with time $t$ in such an SCVP system
is governed by the following kinetic differential equation\cite{Wulkow1997}

\begin{equation}
\frac{dP_{m}(x)}{dt}=R\Big[\frac{1}{2}\sum_{n=1}^{m-1}mP_{n}(x)
P_{m-n}(x)-\sum_{n=1}^{\infty}(n+m)P_{n}(x)P_{m}(x)\Big]
\end{equation}
where $P_{m}(x)$ denotes the number of polymers of $m$-mer at
conversion $x$,
and $R$ is rate constant for the polymerization.
As usual,
the same rate constants of A* and B* reacting with vinyl group A have been used under the assumption of equal reactivity.

Recalling that the initial conversion of vinyl groups A has been arranged to be zero, then there would be $N(1-x)$ free vinyl groups A in the system at a conversion $x$. According to the reaction mechanism, upon considering the change of free vinyl groups A during the polymerization, one can find that the variation of the conversion $x$ in time $t$ is subject to the equation\cite{Wulkow1997}:

\begin{equation}
\frac{d x}{dt}=R\Big[\sum_{m=1}^{\infty}mP_{m}(x=0)\Big](1-x)=RN_{\ast}(1-x)
\end{equation}
where the number of active sites $N_{\ast}$ keeps conservation.
Solving this equation yields
\begin{equation}
x(t)=1-\exp(-RN_{\ast}t)
\end{equation}
This is an explicit relationship between the conversion $x(t)$ and time $t$. The combination of Eqs. (1) and (2) enables one to obtain the following equation
\begin{equation}
\frac{dP_{m}(x)}{dx}=\frac{1}{r}\Big[\frac{m}{2(1-x)}\sum_{n=1}^{m-1}
P_{n}(x)P_{m-n}(x)-(m+\frac{r}{1-x})P_{m}(x)\Big]
\end{equation}
in which $r=N_{\ast}/N$,
denotes the molar ratio of active sites to free vinyl groups at the beginning of the present SCVP system. As $r=1$(this signifies that no polymers is preexisted.), this equation is identical with that given by M\"{u}ller, Yan and Wulkow.\cite{Wulkow1997}

Note that there are $N$ free vinyl groups A at the beginning of the system,
therefore the total number of molecules would be $N(1-x)$ when the
conversion of vinyl groups A is $x$.
This means that $N(1-x)=\sum_{m}P_{m}(x)$.
Furthermore, $P_{m}(x)$ can be rewritten as $P_{m}(x)=N(1-x)c_{m}(x)$ since each molecule possesses only one free vinyl group A,
in which $c_{m}(x)$ denotes the number faction of polymers of $m$-mer at a conversion $x$.
This results in the following identity
\begin{equation}
\sum_{m}c_{m}(x)\equiv 1
\end{equation}
Substituting $P_{m}(x)=N(1-x)c_{m}(x)$ into Eq. (4) yields the differential equation satisfied by $c_{m}(x)$
\begin{equation}
\frac{dc_{m}(x)}{dx}=\frac{m}{r}\Big[\frac{1}{2}\sum_{n=1}^{m-1}
c_{n}(x)c_{m-n}(x)-c_{m}(x)\Big]
\end{equation}

In order to determine $c_{m}(x)$,
a GF denoted by $G(x,\theta)$ can be introduced as follows
\begin{equation}
G(x,\theta)=\sum_{m=1}^{\infty}c_{m}(x)\exp(m\theta)
\end{equation}
where $\theta$ is an auxiliary variable.
Such a GF signifies that $c_{m}(x)$ can be obtained by expanding $G(x,\theta)$
in the series of $\exp(\theta)$ provided that the GF is explicitly given.
It is therefore necessary to derive an explicit expression of $G(x,\theta)$
for further investigations.
To this end, through Eqs. (6) and (7),
one can obtain a partial differential equation satisfied by $G(x,\theta)$, namely
\begin{equation}
[1-G(x,\theta)]\frac{\partial}{\partial \theta}G(x,\theta)+r\frac{\partial}{\partial x}G(x,\theta)=0
\end{equation}
It is obvious that through the initial conditions of the system,
solving this equation will enable us to obtain $G(x,\theta)$,
and hence $c_{m}(x)$ and $P_{m}(x)$.

The partial differential equation given by Eq. (8) is known as Burgers equation\cite{Burgers1974,Logan2008},
and the corresponding characteristics of $G(x,\theta)$ through the point $(0,\phi)$ on the
$x$-$\theta$ plane satisfies the following equation
\begin{equation}
\theta-\phi=\frac{1}{r}[1-G(0,\phi)]x
\end{equation}
where $\phi$ is a function of $x$ and $\theta$ by writing $\phi(x,\theta)$ to be determined.
Clearly, the term $\frac{1}{r}[1-G(0,\phi)]$ plays the role of slope of the characteristic line,
and thus we have $G(x,\theta)=G[0,\phi(x,\theta)]$ from Eq. (8).
This signifies that $G(x,\theta)$ can be derived once the dependence of $\phi(x,\theta)$ on variables $x$ and $\theta$ is figured out.
Making use of $G(x,\theta)=G[0,\phi(x,\theta)]$ together with Eqs. (7) and (9), we have
\begin{equation}
G(x,\theta)=
\sum_{m=1}^{\infty}c_{m}(0)\exp(m\theta)\exp\{-\frac{mx}{r}[1-G(x,\theta)]\}
\end{equation}
From the definition of the GF given in Eq. (7),
the explicit expressions of $G(x,\theta)$ and $c_{m}(x)$ could be directly calculated through the initial conditions given by $c_{m}(0)$.

The derivation of $G(x,\theta)$ is specially useful for the evaluation of average polymeric quantities of the system.
For instance, upon introducing the $k$th ($k=0,1,2,...$) polymer moments
$M_{k}(x)$ defined by $M_{k}(x)=\sum_{m=1}^{\infty}m^{k}P_{m}(x)$, and then
taking into account $P_{m}(x)=N(1-x)c_{m}(x)$ together with Eq. (7) follows that

\begin{equation}
M_{k}(x)=N(1-x)\frac{\partial^{k}}{\partial \theta^{k}}G(x,\theta)|_{\theta=0}
\end{equation}
where $\frac{\partial^{k}}{\partial \theta^{k}}G(x,\theta)$ denotes the $k$th partial derivatives of $G(x,\theta)$ with respect to the variable $\theta$, and $\frac{\partial^{k}}{\partial \theta^{k}}G(x,\theta)|_{\theta=0}$ denotes its value evaluated at $\theta=0$.

Bear in mind that the mass of an inimer has been used as the unit mass,
the number-, weight-, and z-average molecular weights,
and PI can be calculated from the derived $M_{k}(x)$ by using
\begin{eqnarray}
M_{\mathrm{n}}(x)=\frac{M_{1}(x)}{M_{0}(x)},\,\,\,\, M_{\mathrm{w}}(x)=\frac{M_{2}(x)}{M_{1}(x)},\nonumber\\
M_{\mathrm{z}}(x)=\frac{M_{3}(x)}{M_{2}(x)},\,\,\,\,
\mathrm{PI}(x)=\frac{M_{\mathrm{w}}(x)}{M_{\mathrm{n}}(x)}.
\end{eqnarray}
For the system under study, as a consequence of Eqs. (11) and (12), we have
\begin{eqnarray}
M_{\mathrm{n}}(x)=\frac{1}{1-x}M_{\mathrm{n}}(0),\,\,\,\,
M_{\mathrm{w}}(x)=\frac{1}{(1-x)^{2}}M_{\mathrm{w}}(0),\nonumber\\
M_{\mathrm{z}}(x)\!=\!\frac{M_{\mathrm{z}}(0)}{1-x}\!+\!
\frac{3x M_{\mathrm{w}}(0)}{(1-x)^{2}},\,\,\,\,
\mathrm{PI}(x)\!=\!\frac{1}{1-x}\textrm{PI}(0).
\end{eqnarray}
where $M_{\mathrm{n}}(0)$, $M_{\mathrm{w}}(0)$, $M_{\mathrm{z}}(0)$, and PI($0$) are the number-, weight-, and z-average molecular weights,
and PI at $x=0$, respectively. This equation indicates that $M_{\mathrm{n}}(x)$, $M_{\mathrm{w}}(x)$, $M_{\mathrm{z}}(x)$ and PI($x$) have been associated with their corresponding initial values at $x=0$. In other words, the effect of preexisted polymers on these average polymeric quantities has been explicitly carried out.

Now a question that arises naturally is whether the present GF method is valid.
To demonstrate its validity,
the present GF method would be employed to the batch SCVP system of pure inimers due to M\"{u}ller, Yan and Wulkow\cite{Wulkow1997}.
Assume that there exist $N$ inimers in the initial stage of the system,
then we have $c_{m}(0)=\delta_{m,1}$($\delta_{i,j}$ is the Kronecker
symbol, which is equal to 1 for $i=j$, and 0 otherwise.).
Meanwhile, one can find that $r=1$ and
$G(x,\theta)=G[0,\phi(x,\theta)]=\exp[\phi(x,\theta)]$.
With the help of Eqs. (7) and (10), we have
\begin{equation}
G(x,\theta)\exp[-xG(x,\theta)]=\exp(\theta-x)
\end{equation}
Making use of Eqs.(11) and differentiating both sides of this equation with respect to $\theta$,
the polymer moments $M_{0}(x)=N(1-x)$,
$M_{1}(x)=N$, $M_{2}(x)=\frac{N}{(1-x)^{2}}$, and $M_{3}(x)=N\frac{1+2x}{(1-x)^{4}}$ can be easily obtained.
Furthermore, a direct calculation from Eq. (12) leads to
\begin{eqnarray}
M_{\mathrm{n}}(x)&=&\frac{1}{1-x},\,\,\,\,
M_{\mathrm{w}}(x)=\frac{1}{(1-x)^{2}},\nonumber\\
M_{\mathrm{z}}(x)&=&\frac{1+2x}{(1-x)^{2}},\,\,\,\,
\textrm{PI}(x)=\frac{1}{1-x}.
\end{eqnarray}
Alternatively, these results can be directly derived by substituting $M_{\mathrm{n}}(0)$=$M_{\mathrm{w}}(0)$
=$M_{\mathrm{z}}(0)$=1 into Eq. (13).
This signifies that one can obtain these results in terms of the present GF instead of using the size distribution of polymers,
and all we need is merely the initial conditions of the SCVP system.

More interestingly, $G(x,\theta)$ can also be employed to derive the size
distribution of polymers by starting with Eq. (14) together with
the theorem of Lagrange inversion (A detailed calculations can be found in the Appendix A)\cite{Stanley1999}.
For brevity,
here we only give the final expression of $G(x,\theta)$
\begin{equation}
G(x,\theta)=\sum_{m=1}\omega_{m}x^{m-1}\exp(-mx)\exp(m\theta)
\end{equation}
where $\omega_{m}=\frac{m^{m-1}}{m!}$ is rigorously derived from the residue theorem\cite{Kantorovich2016}.
Such an explicit expression of $G(x,\theta)$ means that $c_{m}(x)=\omega_{m}x^{m-1}\exp(-mx)$, and thus the size distribution $P_{m}(x)$ can be given by
\begin{equation}
P_{m}(x)=N(1-x)\omega_{m}x^{m-1}\exp(-mx)
\end{equation}

All the results presented in Eqs. (15) and (17) are identical
with those obtained in the batch SCVP system\cite{Wulkow1997},
thereby validating the present GF method.
Furthermore,
as a direct generalization,
the present GF method will be employed  to discuss the effect of the semi-batch
process on the average polymeric quantities of hyperbranched polymers.
This can be carried out by considering the semi-batch process as a series of multi-step processes of feeding and polymerization because any two nearest neighbour steps have been essentially solved by the present GF method.

\section{A Unified Treatment on the Semi-batch SCVP System}

For a polymerization system under the semi-batch mode, all the feeding operations are usually performed according to a predetermined schedule.
Therefore the whole polymerization is exactly a multi-step process.
This means that, at the beginning of each polymerization step, there always exists some preexisted polymers resulting either from a previous polymerization or from the present feeding, or from both them.
As done in Sec. 2, these discrete steps can be further investigated one by one with the aid of the present GF method.
Throughout the work,
each feeding operation was considered to be instantaneous such that the response of the system during such an infinitesimal time interval can be neglected.

We assume that only inimers are initially loaded into the reactor, which is named the zeroth feeding, for convenience.
Subsequently, the first polymerization step begins until the conversion arrives at a predetermined conversion $x_{1}$ (or equivalently, at a predetermined reaction time.), and then the first feeding is performed.
After the first feeding, the second polymerization begins until the conversion arrives at $x_{2}$.
The rest steps are repeated in a similar manner until the last step polymerization is completed.

In this way,
one can find that a polymerization and a feeding are always involved in each step.
For clarity, a sequence of feedings and reactions in an SCVP system can be schematically illustrated in Fig.1, where the semi-batch process was considered to be an $L$-step process. This indicates that the semi-batch SCVP system has been divided into an $L$-step polymerization process,
and each step can be studied as we have presented in Sec. 2.

\begin{figure}
\centering
\includegraphics[width=3.8in] {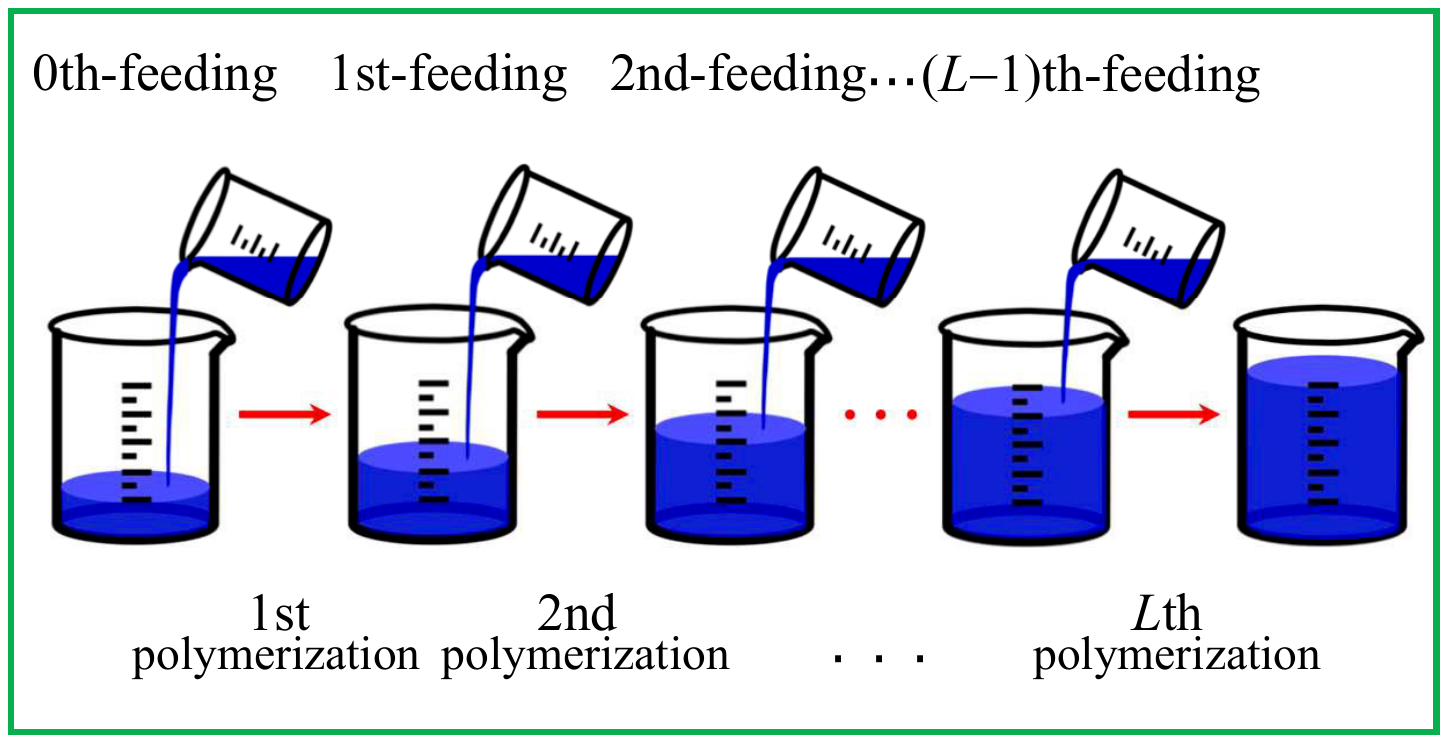}
\caption{Schematic illustration of the semi-batch SCVP system, which is thought of as an $L$-step processes, and each step is comprised of polymerization and feeding.}
\label{fig 1}
\end{figure}

Now the semi-batch SCVP system would be investigated as an application of the present GF method,
which enables us to treat the situation that allows polymers of various sizes to be added into the reactor in each feeding.
From the results given in Sec. 2,
it has been found that these average polymeric quantities associated with two nearest neighbor steps are closely related to each other.
As such,
$M_{\mathrm{n}}$, $M_{\mathrm{w}}$, $M_{\mathrm{z}}$, and PI could be readily derived as long as the feeding details are specified.

\subsection{The GF method for successive feedings of polymers}
As schemed in Fig.1,
for any two nearest neighbor steps, for example, the ($i-1$)th and $i$th steps,
the initial conditions of the $i$th step polymerization depend explicitly on the number of polymers coming from the ($i-1$)th polymerization and feeding.
This indicates that the $i$th polymerization must be affected by the polymerization and feeding in the ($i-1$)th step,
so do the relevant average polymeric quantities.

Note that the ($i-1$)th feeding was performed when the conversion of vinyl groups A was $x_{i-1}$,
and after this feeding the $i$th step polymerization takes place.
Let $P_{m}^{(i)}(x_{i}=0)$ be the number of polymers of $m$-mer
($m=1,2,3,...$) at the beginning of the $i$th step polymerization, which can be expressed as
\begin{eqnarray}
P_{m}^{(i)}(x_{i}=0)=P_{m}(x_{i-1})+F_{m}^{(i-1)}
\end{eqnarray}
where $P_{m}(x_{i-1})$ and $F_{m}^{(i-1)}$ denote the number of polymers of $m$-mer due to the ($i-1$)th polymerization and feeding, respectively.
Obviously, now the total number of active sites is $\sum_{m}mP_{m}^{(i)}(x_{i}=0)$, while the number of vinyl groups A is $N^{(i)}=\sum_{m}P_{m}^{(i)}(x_{i}=0)$.
This is the initial condition of the $i$th step polymerization.

In analogy with Eqs. (9) and (10), the most important two equations resulting from the GF method can be found
\begin{eqnarray}
r_{i}[\theta-\phi(x_{i},\theta)]=[1-G(x_{i},\theta)]x_{i},\nonumber\\
G(x_{i},\theta)=\sum_{m=1}^{\infty}c_{m}^{(i)}(0)\exp[m\phi(x_{i},\theta)]
\end{eqnarray}
where $r_{i}=\sum_{m}mP_{m}^{(i)}(x_{i}=0)/N^{(i)}$, is the molar ratio of active sites to free vinyl groups in the $i$th step,
$x_{i}$ denotes the conversion of vinyl groups A,
and $c_{m}^{(i)}(0)$ is the number fraction of polymers of $m$-mer at $x_{i}=0$.
In obtaining Eq. (19), the equation $G(x_{i},\theta)=G[0,\phi(x_{i},\theta)]$ has been used.

Making use of Eq. (19),
the $k$th polymer moment in the $i$th step polymerization defined by
$M_{k}(x_{i})=\sum_{m=1}^{\infty}m^{k}P_{m}(x_{i})$ can be given by

\begin{equation}
M_{k}(x_{i})=N^{(i)}(1-x_{i})\frac{\partial^{k}}{\partial \theta^{k}}G(x_{i},\theta)|_{\theta=0}
\end{equation}
Likewise, $M_{\mathrm{n}}(x_{i})$, $M_{\mathrm{w}}(x_{i})$, $M_{\mathrm{z}}(x_{i})$, and PI($x_{i}$) in the $i$th step polymerization can be explicitly derived through the $k$th polymer moments evaluated by this equation.
More specifically, as a consequence of the present GF method,
the following recursion formula (A detailed derivation has been left in the Appendix B) can be obtained
\begin{eqnarray}
M_{\mathrm{n}}(x_{i+1})&=&\frac{1}{(1-x_{i+1})}
\frac{f_{1,0}^{(i)}+M_{\mathrm{n}}(x_{i})}{1+f_{0,0}^{(i)}},\nonumber\\
M_{\mathrm{w}}(x_{i+1})&=&\frac{1}{(1-x_{i+1})^{2}} \frac{f_{2,1}^{(i)}+M_{\mathrm{w}}(x_{i})}{1+f_{1,1}^{(i)}},\nonumber\\
M_{\mathrm{z}}(x_{i+1})&=&\frac{f_{3,2}^{(i)}+
M_{\mathrm{z}}(x_{i})}{(1-x_{i+1})(1+f_{2,2}^{(i)})}+\frac{3x_{i+1}
M_{\mathrm{w}}(x_{i})}{(1-x_{i+1})^{2}},\nonumber\\
\mathrm{PI}(x_{i+1})&=&\frac{1+f_{0,0}^{(i)}}{(1-x_{i+1})}
\frac{[f_{2,1}^{(i)}/M_{\mathrm{n}}(x_{i})]+\mathrm{PI}(x_{i})}{[1+f_{1,1}^{(i)}]^{2}},
\end{eqnarray}
where the quantity $f_{k,k'}^{(i)}$ is defined by

\begin{equation}
f_{k,k'}^{(i)}=\frac{\sum_{m}m^{k}F_{m}^{(i)}}{\sum_{m}m^{k'}P_{m}(x_{i})},\,\,\, (k,k'=0,1,2,3,...)
\end{equation}
which denotes the ratio of the $k$th polymeric moment due to feeding
to the $k'$th polymeric moment due to polymerization in the $i$th step.

It should be stressed that the quantity $f_{k,k'}^{(i)}$ depends not only on the number of polymers from the $i$th feeding, but also on those from the $i$th polymerization. Of course, the removal of polymers can be taken into account by stating in Eqs. (18) and (22) the degree of polymerization and the number of those removed polymers. This indicates that $f_{k,k'}^{(i)}$ is closely related to the feeding details such as species and amount of those fed polymers.
For example,
$f_{0,0}^{(i)}$ measures the molar ratio of the two types of polymers
in the $i$th step,
while $f_{1,1}^{(i)}$ measures the corresponding weight ratio.
As such,
$f_{k,k'}^{(i)}$ plays a key role in evaluating average polymer quantities in the $i$th step.
In view of the relevance of two nearest neighbour steps,
$f_{k,k'}^{(i)}$ becomes the most important factor that determines the average properties of resultant polymers.

It can be seen from Eq. (21),
$M_{\mathrm{n}}$, $M_{\mathrm{w}}$, $M_{\mathrm{z}}$,
and PI involved in two nearest neighbor steps have been connected with each other.
In other words,
one can find how an average polymeric quantity in the present step is affected by previous steps in terms of the recursion formula.
In so doing,
we can conclude that these average polymeric quantities have related to all previous steps,
and therefore they can be calculated provided that the feeding details and conversions in each step are specified.
Furthermore,
the effect of the semi-batch process on the SCVP system can be analyzed with the help of Eqs. (21) and (22).
As an illustration, we will devote ourself to the most frequently encountered case that only inimers are fed into in each step, which is of particular importance in practice.
Specifically, two typical feeding ways would be investigated to calculate $M_{\mathrm{n}}$ and $M_{\mathrm{w}}$, as shown below.

\subsection{Case I: Feeding same amount of inimers in each step}
The first feeding way corresponds to the case that the number of fed inimers in each step keeps always the same as each other, which is widely employed in practice and would be referred to as the case I.
This simply means that the number of the fed inimers is equal to that loaded initially (the zero-th feeding in Fig.1).
More precisely, we have for $i\geq 1$

\begin{eqnarray}
\sum_{m}m^{k}F_{m}^{(i)}=\sum_{m}m^{k}F_{m}^{(i)}\delta_{m,1}=F_{1}^{(i)}
\end{eqnarray}
where $\sum_{m}mF_{m}^{(i)}$ also denotes the weight of the fed inimers in the $i$th step since the mass of an inimer has been taken as the unit mass.

Clearly, feeding the same number of inimers in each step
signifies that $f_{1,1}^{(i)}$ satisfies the criteria $f_{1,1}^{(i)}=\frac{1}{i}$.
This results in the following relationships
\begin{eqnarray}
f_{2,1}^{(i)}=f_{1,1}^{(i)}=\frac{1}{i},\,\,\,
f_{1,0}^{(i)}=f_{0,0}^{(i)}=\frac{1}{i}M_{\mathrm{n}}(x_{i})
\end{eqnarray}
where $M_{\mathrm{n}}(x_{i})=\frac{M_{1}(x_{i})}{M_{0}(x_{i})}$, denotes the number-average molecular weight in the $i$th step polymerization with the conversion $x_{i}$.
According to such a feeding operation,
if an SCVP system is eventually terminated at the $L$th polymerization with the conversion $x_{L}$,
from Eqs. (21) and (24) we have
\begin{eqnarray}
\frac{1}{M_{\mathrm{n}}(x_{L})}=
\frac{1}{L}\sum_{j=1}^{L}\big[\prod_{i=j}^{L}(1-x_{i})\big],\nonumber\\
M_{\mathrm{w}}(x_{L})=
\frac{1}{L}\sum_{j=1}^{L}\big[\prod_{i=j}^{L}\frac{1}{(1-x_{i})^{2}}\big].
\end{eqnarray}
In obtaining these,
$M_{\mathrm{n}}(x_{1})=\frac{1}{1-x_{1}}$ and $M_{\mathrm{w}}(x_{1})=\frac{1}{(1-x_{1})^{2}}$ have been used.

The above results manifest that the average properties of resultant polymers
would be influenced by all the previous polymerization steps.
They can be evaluated in a straightforward manner when the conversions of vinyl groups A and feeding details in each step are specified. Namely, for the Case I, one needs $x_{i}$ ($i=1, 2, ...L$) to evaluate $M_{\mathrm{n}}$, $M_{\mathrm{w}}$, $M_{\mathrm{z}}$, and PI.
Furthermore, the effect of the semi-batch process on the SCVP system
under this feeding way can be found.

As an illustration, $M_{\mathrm{w}}$ and
PI for the semi-batch SCVP system under the Case I have been calculated through Eq. (25), as presented in panels (a) and (b) of Fig.2, respectively.
To reveal the effect of the semi-batch mode, the results of $M_{\mathrm{w}}$ and PI under the batch mode are also given.
In our calculations, the conversions $x_{L}$ for different $L$ have been listed in Table 1, in which the final conversions of vinyl groups in the batch and semi-batch modes have been designed to be approximately equal to each other.
Meanwhile, the conversion at each step has been mapped onto the corresponding conversion to compare with the batch process.

\begin{table}
  \centering
  \caption{Conversions at different $L$ for calculating $M_{\mathrm{w}}$ and PI under the Case I.}
  \label{table 1}
  \begin{tabular}{ll}
  \toprule
  Case I & conversions \\
  \midrule
  $L$=2 & $x_{1}=x_{2}=0.85$ \\
  $L$=3 & $x_{1}=x_{2}=x_{3}=0.83$ \\
  $L$=4 & $x_{i}=0.75$, ($i$=1, 2,..., 4) \\
  $L$=5 & $x_{i}=0.70$, ($i$=1, 2,..., 5) \\
  \bottomrule
  \end{tabular}
\end{table}

\begin{figure}
\centering
\includegraphics[height=7.2cm]{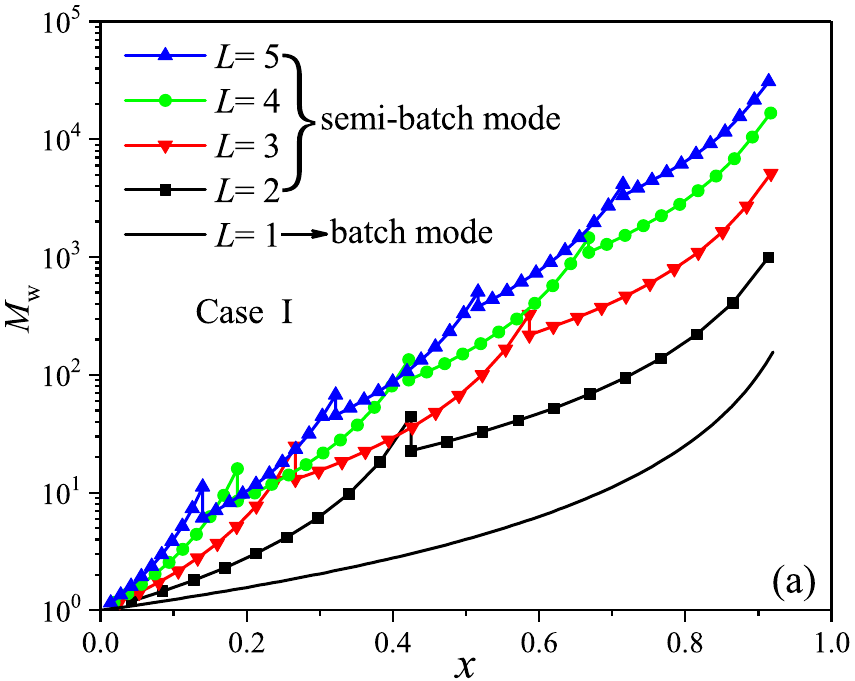}
\includegraphics[height=7cm]{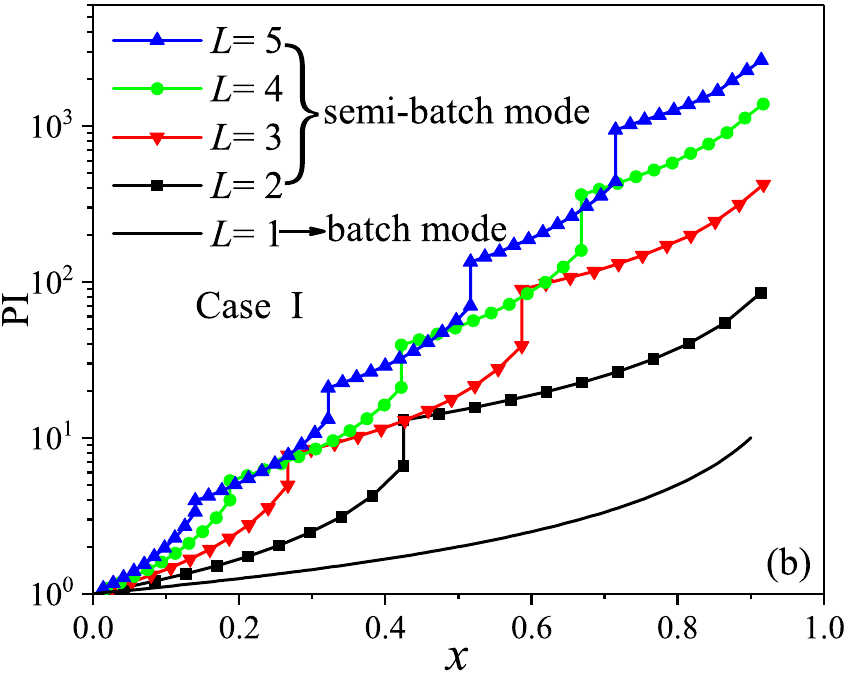}
\caption{$M_{\mathrm{w}}$ in panel (a) and PI in panel (b) for the semi-batch SCVP system under the Case I, where conversions at different $L$ are specified in Table 1, and where the lines are drawn only to guide the eye.}
\label{fig 2}
\end{figure}

Seen from the Fig.2,
we can find the variations of $M_{\mathrm{w}}$ and PI with the increasing conversion at different $L$.
The feeding operations can be obviously observed through the changes in $M_{\mathrm{w}}$ and PI since each feeding the inimers always gives rise to an obvious decrease in $M_{\mathrm{w}}$ and an obvious increase in PI.
Compared with that in the batch mode, $M_{\mathrm{w}}$ and PI in the semi-batch mode can even increase by orders of magnitude.
Because both $M_{\mathrm{w}}$ and PI of polymers play a central role in various fields,
one can take advantage of such an ability of the semi-batch process to prepare hyperbranched polymers with desired $M_{\mathrm{w}}$ and PI.

\subsection{Case II: Feeding inimers to maintain the same number of free vinyl groups in each step}
The second case is that feeding inimers is to maintain a fixed initial concentration of free vinyl groups A,
which is referred to as the case II.
This kind of feeding implies that the consumption fraction of vinyl groups
would be supplied by the feeding operation.
In other words,
the number of vinyl groups at the beginning of each polymerization step always remains a constant.
In a sense, this indicates a steady rate for a polymerization, i.e.,
\begin{eqnarray}
\sum_{m}m^{k}F_{m}^{(i)}\delta_{m,1}=F_{1}^{(i)}=N^{(i)}x_{i}
\end{eqnarray}
where $N^{(i)}$ denotes the initial number of molecules in the $i$th step polymerization, and $N^{(i)}x_{i}$ represents the number of inimers to be supplied by the $i$th feeding.

Given that the number of inimers at the initial stage is $N^{(0)}$,
thus one can find that $N^{(i)}=N^{(0)}$ ($i=1,2,...L$) holds true for all the polymerization steps.
Through these analysis, we obtain
\begin{eqnarray}
f_{1,0}^{(i)}=f_{0,0}^{(i)}=\frac{x_{i}}{1-x_{i}},\nonumber\\
f_{2,1}^{(i)}=f_{1,1}^{(i)}=\frac{x_{i}}{\sum_{j=0}^{i}x_{j}}
\end{eqnarray}
in which we have arranged $x_{0}\equiv 1$, for convenience.
If the SCVP system in eventually terminated at the $L$th polymerization with conversion $x_{L}$,
making use of Eqs. (21) and (27) leads to
\begin{eqnarray}
M_{\mathrm{n}}(x_{L})&=&
\frac{1}{1-x_{L}}\sum_{i=0}^{L-1}x_{i},\nonumber\\
M_{\mathrm{w}}(x_{L})&=&
\frac{1}{\sum_{i=0}^{L-1}x_{i}}\sum_{j=1}^{L}\big[\frac{x_{j-1}}{\prod_{i=j}^{L}(1-x_{i})^{2}}\big]
\end{eqnarray}
In obtaining these equations we have used the results
$M_{\mathrm{n}}(x_{1})=\frac{1}{1-x_{1}}$ and $M_{\mathrm{w}}(x_{1})=\frac{1}{(1-x_{1})^{2}}$.

\begin{table}
  \centering
  \caption{Conversions at different $L$ for calculating $M_{\mathrm{w}}$ and PI under the Case II.}
  \label{table 2}
  \begin{tabular}{ll}
  \toprule
  Case II & conversions \\
  \midrule
  $L$=2 & $x_{1}=x_{2}=0.90$ \\
  $L$=3 & $x_{1}=x_{2}=x_{3}=0.90$ \\
  $L$=4 & $x_{i}=0.90$, ($i$=1, 2,..., 4) \\
  $L$=5 & $x_{i}=0.90$, ($i$=1, 2,..., 5) \\
  \bottomrule
  \end{tabular}
\end{table}

Through Eq. (28), $M_{\mathrm{n}}$, $M_{\mathrm{w}}$, and PI can be evaluated whenever the conversions and feeding details in each step are specified.
It is also easy to find that all the previous steps can result in an apparent effect on the average properties of resultant polymers.
Likewise, for the semi-batch SCVP system under the case II with details given by Table 2, $M_{\mathrm{w}}$ and PI have been carried out, as presented in panels (a) and (b) of Fig.3, respectively.

\begin{figure}
\centering
\includegraphics[height=7cm]{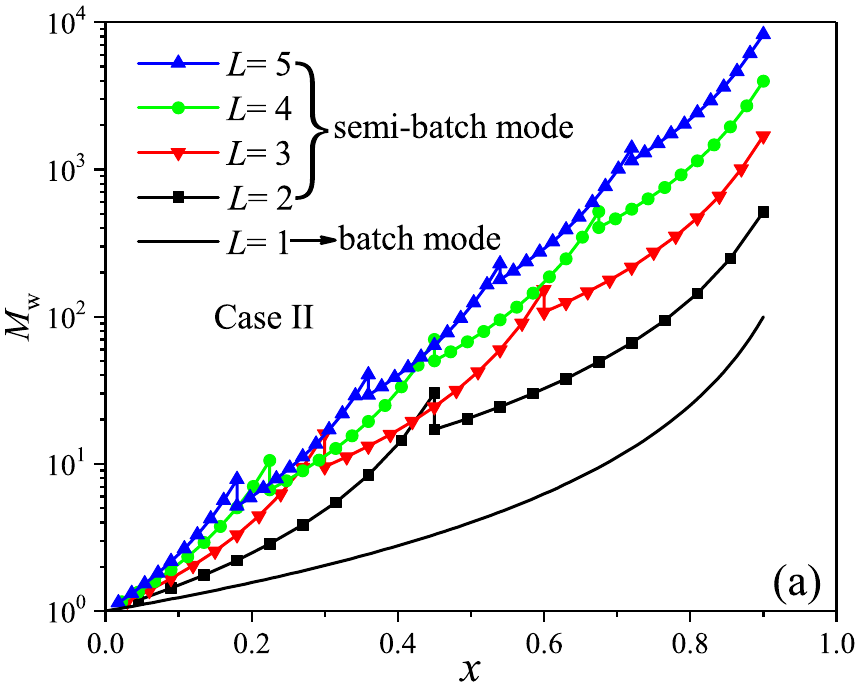}
\includegraphics[height=7cm]{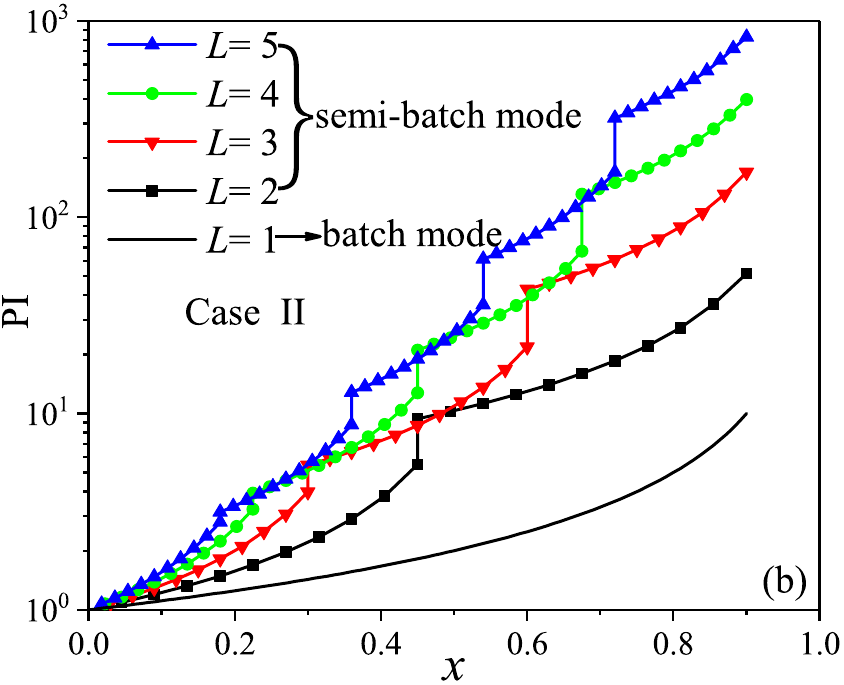}
\caption{$M_{\mathrm{w}}$ in panel (a) and PI in panel (b) for the semi-batch SCVP system under the Case II, where conversions at different $L$ are specified in Table 2, where the lines are drawn only to guide the eye.}
\label{fig 3}
\end{figure}

Compared with that in the batch mode,
$M_{\mathrm{w}}$ and PI in this case also increase by orders of magnitude, as illustrated.
Nevertheless, the difference of the average polymeric quantities between the two cases is rather obvious, as shown in Figs.2 and 3, respectively.
Thus both the two feeding ways can give rise to a significant increase in $M_{\mathrm{w}}$ and $PI$. However, the dependence of final average polymeric quantities on conversions of each step is obviously different from each other.
This implies a significant effect of the distinct feeding ways on the average properties of hyperbranched polymers.

Note that for the present semi-batch SCVP system,
the problem of exotherm could be solved because its characteristic operation of multiple feedings.
In practice,
depending on the specific reaction conditions,
various species and fed amount can be well designed to control the polymerization.
In so doing,
the semi-batch mode becomes compatible with the safe requirement.
This is, of course, particularly important for applying SCVP method to prepare hyperbranched polymers in industry and chemical engineering.

\section{ Monte Carlo Simulation for a Semi-batch SCVP System}

In this section, an MC simulation on the semi-batch SCVP system would be performed for the case that only inimers are fed into the reactor. In particular,
the simulation results for Cases I and II discussed in subsections 3.1 and 3.2 would be compared with the corresponding analytical results.
For a stochastic chemical reaction, a set of master equations can be constructed on the basis of the reaction mechanism\cite{Yang1993,Gillespie2007}.
In the present SCVP system, the change in size distribution of polymers is subject to the differential kinetic equations given in Eq. (1).
Therefore the evolution of MC simulation would be governed by the corresponding master equation.
For easy of presentation,
here we only outline how to use MC method to simulate a semi-batch SCVP system.

Assume that the system consisting of treelike molecules with size distribution ${P_{n}(t)}$ at a given time $t$, and after a time interval $t_{\mathrm{w}}$ a reaction takes place.
As a consequence,
the system evolves into a new state characterized by a renewed size distribution ${P_{n}(t+t_{\mathrm{w}})}$ of polymers at the time $t+t_{\mathrm{w}}$.
The quantity $t_{\mathrm{w}}$ is usually called the waiting time since it denotes a time interval between two successive reactions.
It is therefore important for an MC simulation to determine both the waiting time $t_{\mathrm{w}}$ and reaction type(what species are involved in each reaction).

According to the Eqs. (1) and (2),
if $\Omega_{n}(t)$ is the possible reaction rate of forming treelike polymers of $n$-mer at time $t$, we have
\begin{eqnarray}
\Omega_{n}(t)=\frac{R}{2}\sum_{i=1}^{n-1}nP_{i}(t)[P_{n-i}(t)-\delta_{i,n-i}]
\end{eqnarray}
where the term with Kronecker symbol $\delta_{i,j}$ is to avoid repeated counts of reaction ways.
The total possible reaction rate as a function of $t$ over all possible $n$ can be written as $\Omega(t)=\Sigma_{n}\Omega_{n}(t)$.
In so doing,
the waiting time, $t_{\mathrm{w}}$,
can be determined by a random number $r_{1}$ through the relationship\cite{Yang1993,Gillespie2007}
\begin{equation}
t_{\mathrm{w}}=\Omega(t)^{-1}\ln r_{1}^{-1}
\end{equation}
This indicates that $t_{\mathrm{w}}$ is a stochastic quantity rather than a constant.
Subsequently, $q_{n}(t)$,
the probability of generating treelike molecules of $n$-mer can be expressed as

\begin{equation}
q_{n}(t)=\frac{\Omega_{n}(t)}{\Omega(t)},\,\,\,(n=2,3,...)
\end{equation}

The simulation can be performed by a simple sampling over all reaction events because of the identity of $\sum_{n}q_{n}(t)\equiv 1$.
In other words,
one can sample in the corresponding probability space such that the reaction type associated with a degree of polymerization (index $n$) could be determined.
In such a routine process, two more random numbers would be involved, and the sampling method is available in the literatures.\cite{Yang1993,Gillespie2007}
In this way,
the state characterized by the size distribution of polymers ${P_{n}(t)}$(equivalently, ${P_{n}(x)}$) should be renewed after each reaction and feeding,
whereby the relevant polymer quantities can be evaluated.

To testify the programm,
we have firstly implemented 100 simulations on the batch SCVP system, where the number of inimers in each was fixed as $N=10^{5}$.
As a consequence of the simulation, $M_{\mathrm{n}}$ and $M_{\mathrm{w}}$ have been presented in Fig.4, in which $P_{n}(x)$ at $x=0.97$ is also illustrated in the insert. Meanwhile, a comparison with the corresponding analytic results has been made.
Clearly, the simulation results of $P_{n}$ against $n$, $M_{\mathrm{n}}$ and $M_{\mathrm{w}}$ against $x$ agree well with their analytical results,
thereby validating the present MC simulation method.
This enables us to simulate a semi-batch SCVP system.

\begin{figure}
\centering
\includegraphics[width=3.5in]{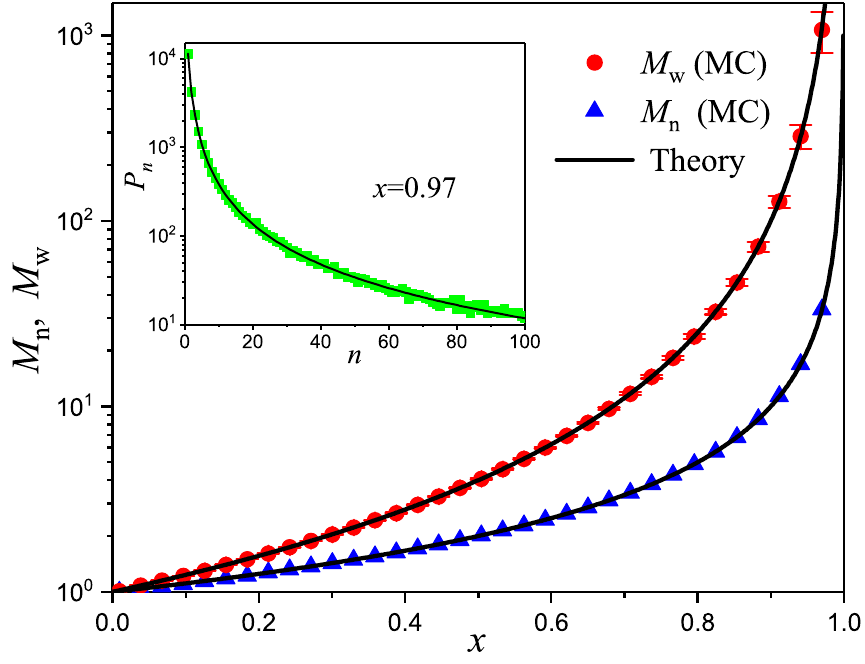}
\caption{ $M_{\mathrm{w}}$(with error bar) and $M_{\mathrm{n}}$ against $x$, and the size distribution $P_{n}(x=0.97)$ against $n$ (insert) for the batch SCVP system, where the lines and symbols denote the theoretical and simulation results, respectively.}
\label{fig 4}
\end{figure}

\begin{figure*}
\centering
\includegraphics[width=10cm]{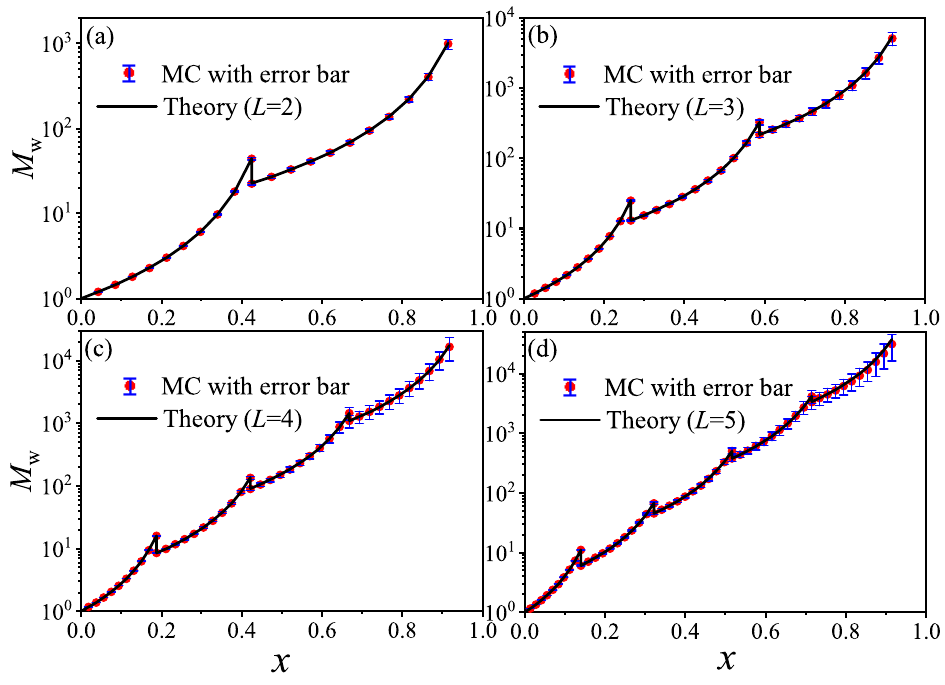}
\caption{$M_{\mathrm{w}}$ for the semi-batch SCVP system in the Case I, where lines and symbols denote the results from the GF method and MC simulations, respectively.}
\label{fig 5}
\end{figure*}

\begin{figure*}
\centering
\includegraphics[width=10cm]{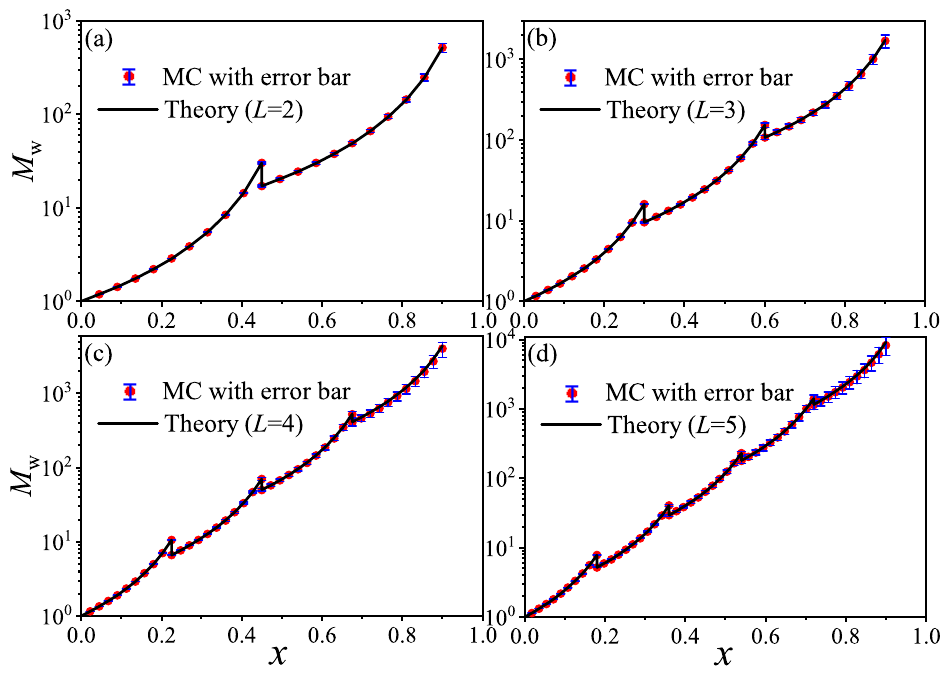}
\caption{$M_{\mathrm{w}}$ for the semi-batch SCVP system in the Case II, where lines and symbols denote the results from the GF method and MC simulation, respectively.}
\label{fig 6}
\end{figure*}

In accordance with the above method,
we have implemented 100 simulations on the semi-batch SCVP system for both Case I and Case II.
In each simulation,
the total number of inimers is fixed as $N=10^{6}$,
while the number of the fed inimers at each step is determined by the conversions specified in the Tables 1 and 2.
For brevity, we only take the simulation results of $M_{\mathrm{w}}$ as an example to compare with the corresponding theoretical results, as shown in
Fig.5 and Fig.6, respectively.

An excellent agreement between the theoretical and simulation results can be observed.
At a fixed $L$,
there was an obvious increase in $M_{\mathrm{w}}$ whenever a polymerization step is finished.
Seen from panels (a) to (d) in Fig.5,
it can be found that $M_{\mathrm{w}}$ increase even by orders of magnitude with the increase in $L$, so does in Fig.6.
This is a general tendency of the present semi-batch SCVP system.
Comparing Fig.5 with Fig.6, one can find that the increment of $M_{\mathrm{w}}$ in the Case I is always larger than that in the Case II.
Such a tendency depends on the predetermined schedule given in Tables 1 and 2.

The above simulation results manifested that the present GF method can be employed to evaluate $M_{\mathrm{n}}$, $M_{\mathrm{w}}$, $M_{\mathrm{z}}$, and PI of polymers by a semi-batch SCVP system. Experimently, one can optimize feeding details to prepare hyperbranched polymers with desired average properties. Consequently, one can choose feeding ways, specify the number of batches, and predetermine conversions at each step. On the basis of these information, some average polymeric quantities of interest can be obtained by Eq. (21) in a straightforward manner.

\section{ Conclusions}

In summary,
we have presented a GF method to investigate the SCVP system under the semi-batch mode by treating it as an $L$-step process.
In each step the feeding allows various polymers produced from another SCVP system to be added into the reactor.
In this way,
the proposed GF method can be employed to calculate some average polymeric quantities such as $M_{\mathrm{n}}$, $M_{\mathrm{w}}$, $M_{\mathrm{z}}$, and PI when the predetermined details on the polymerization are specified.
The relevant results were confirmed by the MC simulation, and therefore a unified treatment on the semi-batch SCVP system of pure inimers has been made.

Clearly, given the predetermined schedule that contains the number of batches $L$ and the feeding details(feeding way, species and their amounts, and conversions in each step, and so forth),
some average properties of the semi-batch SCVP system can be easily evaluated by the present GF method.
In view of the fact that the reactant concentrations in the semi-batch process are controllable, the drifts of molecular weight distribution can be substantially inhibited.
Meanwhile, this also reduce the risk of the maximum exotherm, thereby maintaining a stable polymerization condition.

Although the removal of various polymers from the reactor is not yet involved in this work, it is not difficult to generalize the GF method to take into account such operations.
This can be realized by adding the corresponding terms in Eqs. (18) and (22) by specifying the degree of polymerization and amount of those removed polymers in each step. In so doing, the effect of the removal operation on the SCVP system can be found.

Note that the present semi-batch SCVP system was treated as an $L$-step process, and a further study on the quasi-continuous process can be performed by increasing $L$. In particular, it is even possible that an SCVP system with a continuous feeding can be investigated for a large $L$, where the feeding rate plays an important role.
Meanwhile, the binary SCVP system consisting of inimers and core initiator under the semi-btach mode can be readily studied by the present GF method.
The relevant investigations are underway.

\section*{Acknowledgement}
This work is supported  by the Project for Talent Engineering of Hebei Province under the grant No.A2016015001.
Dr. Gu is gratefully acknowledged the Project for Top Young Talent of Hebei Province and that for general colleges of Hebei Province (No. BJ2017017).

\newpage
\appendix
\section{Derivation of the size distribution from the GF method}

\setcounter{equation}{0}
\renewcommand{\theequation}{A\arabic{equation}}

Here we will derive the size distribution of the batch SCVP system of pure inimers through the present GF method. To proceed, starting from Eq. (14),
one has,
\begin{equation}
G(x,\theta)\exp[-xG(x,\theta)]=\exp(\theta-x)
\end{equation}
Arranging $f(G)$=$G(x,\theta)\exp[-xG(x,\theta)]$ and $X$=$\exp(\theta-x)$, one can find that
\begin{equation}
f(G)=G(x,\theta)\exp[-xG(x,\theta)]=X
\end{equation}

Recalling that the GF is defined by $G(x,\theta)=\sum_{m=1}^{\infty}c_{m}(x)\exp(m\theta)$, then finding the inversion form $G(X)$=$f^{-1}(X)=G(x,\theta)$ is an efficient way to derive $c_{m}(x)$.
For this purpose, we can expand the $G$ as a series of the variable $X$ in terms of the Lagrange inversion theorem\cite{Stanley1999},
\begin{equation}
G(X)=\sum_{m=1}a_{m}X^{m}=\sum_{m}a_{m}\exp[m(\theta-x)]
\end{equation}
where the coefficient $a_{m}$ can be given by the following contour integration around the origin\cite{Kantorovich2016}
\begin{equation}
a_{m}=\frac{1}{2\pi i}\oint\frac{G}{X^{m+1}}dX=\frac{1}{2\pi i}\oint\frac{Gf'(G)}{[f(G)]^{m+1}}dG
\end{equation}
in which $f'(G)$ denotes the first derivative of $f(G)$ with respect to $G$, and $i^{2}=-1$.

Making using of the residue theorem\cite{Kantorovich2016} yields $a_{m}=\frac{m^{m-1}}{m!}x^{m-1}$, and then substituting $a_{m}$ into Eq. (A3) arrives at Eq. (16) in the main text, namely
\begin{equation}
G(x,z)=\sum_{m=1}\omega_{m}x^{m-1}\exp(-mx)\exp(m\theta)
\end{equation}
where $\omega_{m}=\frac{m^{m-1}}{m!}$. This means that $c_{m}(x)=\omega_{m}x^{m-1}\exp(-mx)$, and hence the size distribution of $P_{m}(x)=N(1-x)c_{m}(x)$ can be expressed as
\begin{equation}
P_{m}(x)=N(1-x)\omega_{m}x^{m-1}\exp(-mx)
\end{equation}
which is identical with the result due to M\"{u}ller, Yan, and Wulkow, who derived it by solving the kinetic differential equation.\cite{Wulkow1997}

\section{Derivation of the recursion formula Eq. (21) in the main text}

\renewcommand{\theequation}{B\arabic{equation}}
\setcounter{equation}{0}
Now we derive the recursion formula Eq. (21) in connection with the number-, weight-, and z-average molecular weights, and PI.
Note that after the $(i-1)$th feeding, the $i$th polymerization begins until the conversion of vinyl groups becomes $x_{i}$.
Correspondingly, these quantities become $M_{\mathrm{n}}(x_{i})$, $M_{\mathrm{w}}(x_{i})$, $M_{\mathrm{z}}(x_{i})$, and PI($x_{i}$). Of the particular interest is the relationship between these average polymeric quantities associated with the such successive two steps.

According to the definitions of $M_{\mathrm{n}}(x_{i})$, $M_{\mathrm{w}}(x_{i})$, and $M_{\mathrm{z}}(x_{i})$, these polymeric quantities are closely related to the $k$th polymer moment $M_{k}(x_{i})$ defined by $M_{k}(x_{i})=\sum_{m=1}^{\infty}m^{k}P_{m}(x_{i})$.
Also, $M_{k}(x_{i})$ is associated with $G(x_{i},\theta)$ by Eq. (20),
and thus we have

\begin{equation}
M_{\mathrm{n}}(x_{i})=\frac{\sum_{m=1}^{\infty}mP_{m}(x_{i})}
{\sum_{m=1}^{\infty}P_{m}(x_{i})}=
\frac{G'(x_{i},\theta)}{G(x_{i},\theta)}|_{\theta=0}
\end{equation}
\begin{equation}
M_{\mathrm{w}}(x_{i})=\frac{\sum_{m=1}^{\infty}m^{2}P_{m}(x_{i})}
{\sum_{m=1}^{\infty}mP_{m}(x_{i})}=
\frac{G''(x_{i},\theta)}{G'(x_{i},\theta)}|_{\theta=0}
\end{equation}
\begin{equation}
M_{\mathrm{z}}(x_{i})=\frac{\sum_{m=1}^{\infty}m^{3}P_{m}(x_{i})}
{\sum_{m=1}^{\infty}m^{2}P_{m}(x_{i})}=
\frac{G'''(x_{i},\theta)}{G''(x_{i},\theta)}|_{\theta=0}.
\end{equation}
where $G'(x_{i},\theta)$, $G''(x_{i},\theta)$, and $G'''(x_{i},\theta)$ denote the first, second, and third derivatives of $G(x_{i},\theta)$ with respect to $\theta$, which can be evaluated in terms of Eq. (19), namely,

\begin{equation}
G(x_{i},\theta)=\sum_{m=1}^{\infty}c_{m}^{(i)}(0)\exp[m\phi(x_{i},\theta)]
\end{equation}
\begin{equation}
\phi(x_{i},\theta)=\theta-\frac{x_{i}}{r_{i}}[1-G(x_{i},\theta)],
\end{equation}
where $r_{i}$ is the molar ratio of active sites to free vinyl groups in the $i$th polymerization given by $r_{i}=\frac{\sum_{m}mP_{m}^{(i)}(x_{i}=0)}{\sum_{m}P_{m}^{(i)}(x_{i}=0)}
=\frac{\sum_{m}mc_{m}^{(i)}(0)}{\sum_{m}c_{m}^{(i)}(0)}=\sum_{m}mc_{m}^{(i)}(0)$.
Here the factor $c_{m}^{(i)}(0)$ should be understood as $c_{m}^{(i)}(x_{i}=0)$.

From Eqs. (B4) and (B5), it is easy to find that

\begin{equation}
G'(x_{i},\theta)=
\frac{\sum_{m=1}^{\infty}mc_{m}^{(i)}(0)\exp(m\phi)}
{1-\frac{x_{i}}{r_{i}}
\sum_{m=1}^{\infty}mc_{m}^{(i)}(0)\exp(m\phi)}
\end{equation}
\begin{equation}
G''(x_{i},\theta)\!=\!
\frac{\sum_{m=1}^{\infty}m^{2}c_{m}^{(i)}(0)\exp(m\phi)}
{1-\frac{x_{i}}{r_{i}}
\sum_{m=1}^{\infty}mc_{m}^{(i)}(0)\exp(m\phi)}
[1+\frac{x_{i}}{r_{i}}G'(x_{i},\theta)]^{2}
\end{equation}
\begin{equation}
\begin{split}
G'''(x_{i},\theta)=&
\frac{\sum_{m=1}^{\infty}m^{3}c_{m}^{(i)}(0)\exp(m\phi)
[1+\frac{x_{i}}{r_{i}}G'(x_{i},\theta)]^{3}}
{1-\frac{x_{i}}{r_{i}}
\sum_{m=1}^{\infty}mc_{m}^{(i)}(0)\exp(m\phi)}\\
&+\frac{3x_{i}}{r_{i}}\frac{[G''(x_{i},\theta)]^{2}}
{[1+\frac{x_{i}}{r_{i}}G'(x_{i},\theta)]}.
\end{split}
\end{equation}

As $\theta$ is equal to zero, we have $\phi(x_{i},\theta)=0$ from Eq. (19) because of $G(x_{i},\theta)\equiv 1$. Accordingly, Eqs. (B1)-(B3) becomes

\begin{equation}
M_{\mathrm{n}}(x_{i})=\frac{1}{(1-x_{i})}M_{\mathrm{n}}(x_{i}=0)
\end{equation}
\begin{equation}
M_{\mathrm{w}}(x_{i})=\frac{1}{(1-x_{i})^{2}}M_{\mathrm{w}}(x_{i}=0)
\end{equation}
\begin{equation}
M_{\mathrm{z}}(x_{i})=\frac{M_{\mathrm{z}}(x_{i}=0)}{(1-x_{i})}
+\frac{3x_{i}M_{\mathrm{w}}(x_{i}=0)}{(1-x_{i})^{2}}.
\end{equation}

Clearly, these equations related some average polymeric quantities at conversion $x_{i}$ to those at conversion $x_{i}=0$. In other words, average polymeric quantities are determined by initial reaction conditions of the $i$th polymerization. Note that at the beginning of the $i$th polymerization($x_{i}=0$), $P_{m}^{(i)}=P_{m}(x_{i-1})+F_{m}^{(i-1)}$ holds true from Eq. (18). This leads to the following equations

\begin{equation}
M_{\mathrm{n}}(x_{i}=0)=
\frac{\sum_{m=1}^{\infty}m[P_{m}(x_{i-1})+F_{m}^{(i-1)}]}
{\sum_{m=1}^{\infty}[P_{m}(x_{i-1})+F_{m}^{(i-1)}]}
\end{equation}
\begin{equation}
M_{\mathrm{w}}(x_{i}=0)=
\frac{\sum_{m=1}^{\infty}m^{2}[P_{m}(x_{i-1})+F_{m}^{(i-1)}]}
{\sum_{m=1}^{\infty}m[P_{m}(x_{i-1})+F_{m}^{(i-1)}]}
\end{equation}
\begin{equation}
M_{\mathrm{z}}(x_{i}=0)=
\frac{\sum_{m=1}^{\infty}m^{3}[P_{m}(x_{i-1})+F_{m}^{(i-1)}]}
{\sum_{m=1}^{\infty}m^{2}[P_{m}(x_{i-1})+F_{m}^{(i-1)}]}.
\end{equation}

Furthermore, through a straightforward calculation, we arrive at the following recursion formula

\begin{equation}
M_{\mathrm{n}}(x_{i+1})=\frac{1}{(1-x_{i+1})}
\frac{f_{1,0}^{(i)}+M_{\mathrm{n}}(x_{i})}{1+f_{0,0}^{(i)}}
\end{equation}
\begin{equation}
M_{\mathrm{w}}(x_{i+1})=\frac{1}{(1-x_{i+1})^{2}} \frac{f_{2,1}^{(i)}+M_{\mathrm{w}}(x_{i})}{1+f_{1,1}^{(i)}}
\end{equation}
\begin{equation}
M_{\mathrm{z}}(x_{i+1})=\frac{f_{3,2}^{(i)}+
M_{\mathrm{z}}(x_{i})}{(1-x_{i+1})(1+f_{2,2}^{(i)})}+
\frac{3x_{i+1}M_{\mathrm{w}}(x_{i})}{(1-x_{i+1})^{2}}
\end{equation}
\begin{equation}
\mathrm{PI}(x_{i+1})=\frac{1+f_{0,0}^{(i)}}{(1-x_{i+1})}
\frac{[f_{2,1}^{(i)}/M_{\mathrm{n}}(x_{i})]+\mathrm{PI}(x_{i})}{[1+f_{1,1}^{(i)}]^{2}},
\end{equation}
where the quantity $f_{k,k'}^{(i)}$ ($k,k'=0,1,2,3,...$) defined by Eq. (22) is associated with the feeding details. This is the equation given by Eq. (21) in the main text.

As expected, the effect of the semi-batch mode on average polymeric quantities is obviously reflected by $f_{k,k'}^{(i)}$.
This means that the proposed GF method indeed provides an effective way to tackle the semi-batch SCVP system. This will be helpful to optimize feeding details an predetermine schedule under the semi-batch mode.


\bibliographystyle{elsarticle-num}
\biboptions{sort&compress}
\newpage
\bibliography{Polymer}


%
%
%
\end{document}